\documentclass[reprint, superscriptaddress]{revtex4-2}

\usepackage{tensor}
\usepackage{amsmath}
\usepackage{amssymb}
\usepackage{enumerate}
\usepackage{mathrsfs}
\usepackage{mathtools}
\usepackage{graphicx}
\usepackage{stmaryrd}
\usepackage{tikz-cd}
\usepackage{comment}
\usepackage[compat=1.1.0]{tikz-feynman}
\usepackage{twistor}
\usepackage{graphicx}% Include figure files
\usepackage{dcolumn}% Align table columns on decimal point
\usepackage{bm}% bold math
\usepackage{hyperref}% add hypertext capabilities
\newcommand{\eps}{\epsilon}

\newcommand{\nbar}{\bar\nabla}
\newcommand{\qed}{\hfill\ensuremath{\Box}}

\newtheorem{propn1}{Proposition}
\newtheorem{corol1}{Corollary}[propn1]

\renewcommand{\d}{\mathrm{d}}

\begin{document}

%\preprint{APS/123-QED}

\title{Twistor action for general relativity}

\author{Atul Sharma}
\affiliation{The Mathematical Institute, University of Oxford, OX2 6GG, United Kingdom}
\email{atul.sharma@maths.ox.ac.uk}

\date{\today}

\begin{abstract}
We reformulate Euclidean general relativity without cosmological constant as an action governing the complex structure of twistor space. Extending Penrose's non-linear graviton construction, we find a correspondence between twistor spaces with partially integrable almost complex structures and four-dimensional space-times with off-shell metrics. Using this, we prove that our twistor action reduces to Plebanski's action for general relativity via the Penrose transform. This should lead to new insights into the geometry of graviton scattering as well as to the derivation of computational tools like gravitational MHV rules.
\end{abstract}

\maketitle

%\tableofcontents

\section{Introduction}

Dualities relating space-time field theories and holomorphic theories on twistor space lie at the heart of many remarkable structures in scattering amplitudes. Twistor and ambitwistor strings give rise to worldsheet formulae for all tree-level amplitudes in $\cN=4$ super-Yang-Mills \cite{Witten:2003nn, Berkovits:2004hg, Roiban:2004yf, Geyer:2014fka}. The gauge theory twistor action was originally discovered as an effective action of twistor strings and proved to be equivalent to the Yang-Mills action up to a topological $\theta$-term \cite{Mason:2005zm, Boels:2006ir}. This led to constructive proofs of the Parke-Taylor formula, the maximally helicity violating (MHV) diagram formalism \cite{Cachazo:2004kj, Boels:2007qn, Adamo:2011cb}, the amplitude-Wilson loop duality \cite{Mason:2010yk, Bullimore:2011ni, Adamo:2011pv}, and numerous other correspondences \cite{Adamo:2011dq, Adamo:2011cd, Koster:2014fva, Chicherin:2014uca, Koster:2016ebi, Koster:2016loo, Koster:2016fna}. Recently, it has also yielded the first ever all-multiplicity results on gluon scattering in non-trivial backgrounds \cite{Adamo:2020syc,Adamo:2020yzi}. 

On the other hand, the long-sought twistor action for general relativity (GR) has proven to be much more elusive. A twistor string for gravity was formulated in \cite{Skinner:2013xp} and gave rise to the tree amplitudes of $\cN=8$ supergravity \cite{Hodges:2012ym, Cachazo:2012kg, Cachazo:2012pz}, but it lacked an effective action description. Direct attempts at finding MHV rules for graviton scattering were also carried out in \cite{BjerrumBohr:2005jr}, but broke down at high multiplicity \cite{Bianchi:2008pu}. Meanwhile, twistor actions for conformal gravity \cite{Berkovits:2004jj,Mason:2005zm} and self-dual GR \cite{Wolf:2007tx, Mason:2007ct} were successfully constructed and later expanded to encode leading-order non-self-dual interactions \cite{Mason:2008jy}. These were able to constructively reproduce tree-level graviton MHV amplitudes \cite{Mason:2008jy, Adamo:2013tja, Adamo:2013cra, Adamo:2021bej}, but lacked any manifest equivalence with GR beyond the MHV sector. Further investigations encountered similar roadblocks \cite{Herfray:2016qvg}. 

In this letter, we present a new twistor action that is equivalent to the chiral action for Euclidean GR (without cosmological constant) discussed in \cite{AbouZeid:2005dg}. Our main tool is a novel generalization of Penrose's non-linear graviton construction \cite{Penrose:1976js} that associates certain almost complex structures on twistor spaces to space-times with \emph{off-shell} metrics. Our action also encodes the non-self-dual sector of GR, providing a classical but fully non-linear resolution of the long-standing googly problem of twistor theory \cite{Penrose:2015lla}.  This represents a significant step toward the construction of twistor spaces for non-self-dual solutions of Einstein's equations. 
%Combined with the gauge theory twistor action, this could also provide a novel way to double copy non-self-dual classical solutions in the spirit of \cite{White:2020sfn}. 
Furthermore, it paves a clear way for the derivation of an MHV formalism for gravity by means of its perturbative expansion.

%%%%%%%%%%%%%%%%%%%%%%%
%%%%%%%%%%%%%%%%%%%%%%%

\section{Chiral formulation of GR}\label{sec:pleb}

Let $\cM$ be a four-dimensional manifold with Riemannian metric $g$. We continue to call it ``space-time''. We can introduce a (complex) null tetrad $e^{\al\dal}$ for this metric,
\be\label{ds2}
\d s^2 = \eps_{\al\beta}\,\eps_{\dal\dot\beta}\,e^{\al\dal}\,e^{\beta\dot\beta}\,,
\ee
where $\al=0,1$, $\dal=\dot0,\dot1$ are spinor indices and $\eps_{\al\beta}$, $\eps_{\dal\dot\beta}$ are Levi-Civita symbols. Spinor indices are raised using $\eps^{\al\beta}$, $\eps^{\dal\dot\beta}$ satisfying $\eps^{\al\beta}\eps_{\gamma\beta} = \delta^\al_\gamma$ and $\eps^{\dal\dot\beta}\eps_{\dot\gamma\dot\beta} = \delta^{\dal}_{\dot\gamma}$. Spinor contractions are conventionally denoted by $\la\lambda\,\kappa\ra=\lambda^\al\kappa_\al$, $[\mu\,\rho]=\mu^{\dal}\rho_{\dal}$, etc.

The anti-self-dual (ASD) 2-forms are spanned by
\be
\Sigma^{\al\beta} = \Sigma^{(\al\beta)} = e^{\al\dal}\wedge e^{\beta}{}_{\dal}\,.
\ee
Using these, we work with a version of Plebanski's chiral action for GR espoused in \cite{AbouZeid:2005dg}:
\be\label{plebac}
S[e,\Gamma] = \int_\cM\Sigma^{\al\beta}\wedge(\d\Gamma_{\al\beta}+\kappa^2\,\Gamma_\al{}^\gamma\wedge\Gamma_{\gamma\beta})
\ee
given in terms of the tetrad and auxiliary 1-form fields $\Gamma_{\al\beta} = \Gamma_{(\al\beta)}$, where $\kappa$ is the gravitational coupling. This chiral action is equivalent to the Einstein-Hilbert action up to a topological term. The equation of motion of $\Gamma_{\al\beta}$ sets $\kappa^2\Gamma_{\al\beta}$ to equal the ASD spin connection associated to $g$. The tetrad's equation of motion then implies Ricci-flatness. In the self-dual (SD) limit $\kappa\to0$ of GR, an integration by parts reduces this action to \footnote{We will neglect the boundary term as we are restricting attention to asymptotically flat space-times. Like the topological term, it may need to be reinstated when turning on a cosmological constant.}
\be\label{sdplebac}
S_\text{SD}[e,\Gamma] = \int_\cM\Gamma_{\al\beta}\wedge\d\Sigma^{\al\beta}\,.
\ee
Here, $\Gamma_{\al\beta}$ acts as a Lagrange multiplier and imposes the closure of $\Sigma^{\al\beta}$. In this case, it follows from the structure equation for $\Sigma^{\al\beta}$ that the ASD spin connection is flat and the space-time is self-dual vacuum.

%%%%%%%%%%%%%%%%%%%%%%%
%%%%%%%%%%%%%%%%%%%%%%%

\section{Euclidean twistor theory}

We start by recalling the twistor correspondence for Euclidean signature flat space (see \cite{Woodhouse:1985id,Jiang:2008xw, Adamo:2017qyl} for a review). The twistor space of $\R^4$ is $\PT=\P^3\,\backslash\,\P^1$. This is also the total space of the holomorphic vector bundle $\cO(1)\oplus\cO(1)\to\P^1$. Let $Z^A=(\mu^{\dal},\lambda_\al)$ be homogeneous twistor coordinates, with $\lambda_\al$ denoting coordinates on the base $\P^1$ and $\mu^{\dal}$ up the fibers of $\cO(1)\oplus\cO(1)$. We endow $\PT$ with a reality structure induced by the quaternionic conjugation: $Z^A\mapsto\hat Z^A = (\hat\mu^{\dal},\hat\lambda_\al)$ with
\be
\hat\lambda_\al = (\overline{\lambda_1},-\overline{\lambda_0})\,,\qquad\hat\mu^{\dal} = (\overline{\mu^{\dot1}},-\overline{\mu^{\dot0}})\,.
\ee
The points $x^{\al\dal}\in\R^4$ of flat space are in 1:1 correspondence with projective lines in twistor space that are left invariant by the $\hat{\cdot}$ conjugation:
\be
x^{\al\dal}\quad\longleftrightarrow\quad X\simeq\P^1 :\;\;\mu^{\dal}=x^{\al\dal}\lambda_\al
\ee
that simultaneously satisfy $\hat\mu^{\dal}=x^{\al\dal}\hat\lambda_\al$. This correspondence recovers $\R^4$ as the moduli space of such lines.

If we let $x$ vary, pullback to these real twistor lines provides a diffeomorphism between $\PT$ and the projective spinor bundle of undotted spinors $\PS = \R^4\times\P^1$ with coordinates $(x^{\al\dal},\lambda_\al)$. It is useful to work directly on $\PS$ when building action principles. The $(0,1)$-vector fields determining the twistor complex structure on $\PS$ are spanned by
\be
\dbar_0 = -\la\lambda\,\hat\lambda\ra\,\lambda_\al\,\frac{\p}{\p\hat\lambda_\al}\,,\qquad \dbar_{\dal} = \lambda^\al\,\p_{\al\dal}\,,
\ee
where $\p_{\al\dal}\equiv\p/\p x^{\al\dal}$. Their dual $(0,1)$-forms are
\be
\bar e^0 = \frac{\D\hat\lambda}{\la\lambda\,\hat\lambda\ra^2}\,,\qquad\bar e^{\dal} = \frac{\hat\lambda_\al\,\d x^{\al\dal}}{\la\lambda\,\hat\lambda\ra}\,,
\ee
where $\D\hat\lambda \equiv \la\hat\lambda\,\d\hat\lambda\ra$. We also list convenient bases of $(1,0)$-vector fields and $(1,0)$-forms:
\begin{align}
\p_0 &= \frac{\hat\lambda_\al}{\la\lambda\,\hat\lambda\ra}\,\frac{\p}{\p\lambda_\al}\,,\hspace{-2em}&&\p_{\dal} = -\frac{\hat\lambda^\al\,\p_{\al\dal}}{\la\lambda\,\hat\lambda\ra}\,,\\
e^0 &= \D\lambda\,,&&e^{\dal} = \lambda_\al\,\d x^{\al\dal}\,,
\end{align}
where $\D\lambda \equiv \la\lambda\,\d\lambda\ra$ is the canonical holomorphic top-form on $\P^1$. In terms of these, we can equip $\PT$ with a holomorphic Poisson structure through the bivector
\be\label{I}
I = \eps^{\dal\dot\beta}\,\p_{\dal}\wedge\p_{\dot\beta}
\ee
whose symplectic leaves are the fibers of $\cO(1)\oplus\cO(1)$. 

In the computations below, we also use the fact that exterior derivatives of projective differential forms on $\PS$ with homogeneity $n$ in $\lambda_\al$ and $0$ in $\hat\lambda_\al$ receive corrections from the Chern connection on $\cO(n)\to\P^1$:
\be\label{Chern}
\d_{\PS} \equiv \d = \d_{\bbS} + n\,\frac{\la\hat\lambda\,\d\lambda\ra}{\la\lambda\,\hat\lambda\ra}\wedge\;\;,
\ee
where $\bbS = \R^4\times\C^2$ is the non-projective spinor bundle.

%%%%%%%%%%%%%%%%%%%%%%%
%%%%%%%%%%%%%%%%%%%%%%%

\section{Off-shell non-linear graviton}

\paragraph{Curved twistor spaces.} 
%Penrose's non-linear graviton constructs curved SD space-times using integrable complex structure deformations of $\PT$. We generalize this to also encode off-shell Riemannian metrics by means of only partially integrable deformations.

Instead of working covariantly with the Atiyah-Hitchin-Singer almost complex structure \cite{Atiyah:1978wi,Woodhouse:1985id} like in \cite{Mason:2005zm, Mason:2007ct,Herfray:2016qvg}, we now build a new \emph{local} model of twistor spaces for off-shell curved space-times. Penrose's non-linear graviton \cite{Penrose:1976js, Ward:1990vs} will emerge as a corollary. 

Let $\CPT$ be a manifold that is diffeomorphic to $\PT$ (equivalently $\PS$) and possesses an almost complex structure with Dolbeault operator
\be
\nbar = \dbar + V\,.
\ee 
We assume that, like $\PT$, it has a fibration $\CPT\to\P^1$ that is at least smooth. This lets us use twistor coordinates $Z^A$ as well as spinor bundle coordinates $(x,\lambda)$ as local coordinates on $\CPT$ (when using the latter, we occasionally abuse notation and refer to $\CPT$ by $\PS$ as well). In these coordinates, $\dbar$ is the ``background'' Dolbeault operator on $\PS$,
\be
\dbar = \bar e^0\,\dbar_0 + \bar e^{\dal}\,\dbar_{\dal}\,,
\ee
while $V\in\Omega^{0,1}(\PS,T^{1,0}_{\PS})$ provides a finite deformation
\be
V\equiv V^{\dal}\,\p_{\dal} = \bigl(\bar e^0\,V_0{}^{\dal} + \bar e^{\dot\beta}\,V_{\dot\beta}{}^{\dal}\bigr)\,\p_{\dal}\,.
\ee
Occasionally, we also set $V_0{}^{\dal}\,\p_{\dal}\equiv V_0$ and $V_{\dot\beta}{}^{\dal}\,\p_{\dal}\equiv V_{\dot\beta}$. 
For the deformation to be compatible with the fibration $\CPT\to\P^1$, we have taken $V\,\lrcorner\, e^0=0$.

In what follows, we will also need to assume that $V$ is a hamiltonian vector field with respect to the Poisson bivector $I$ given in \eqref{I}. This leads to a zero-divergence condition on the $(0,1)$-form valued components $V^{\dal}$:
\be\label{divfree}
%\begin{split}
\text{div}\,V\equiv\cL_{\p_{\dal}}V^{\dal} = 0\,.\\
%\implies \p_{\dal}V_0{}^{\dal}+V_{\dal}{}^{\dal} = 0 &= \p_{\dal}V_{\dot\beta}{}^{\dal}\,.
%\end{split}
\ee
We will impose this as a constraint in our action, though one can also solve it in terms of a hamiltonian $h$ \cite{Mason:2007ct}.

In the deformed complex structure, the $(0,1)$-vector fields are spanned by $\dbar_0+V_0$, $\dbar_{\dal}+V_{\dal}$. The associated basis of $(1,0)$-forms on $\CPT$ is
\be
e^0=\D\lambda\,,\quad \theta^{\dal} = e^{\dal} - V^{\dal}\,,
\ee
%\be
%e^0=\D\lambda\,,\quad \theta^{\dal} = e^{\dal} - \bar e^0\,V_0{}^{\dal}-\bar e^{\dot\beta}\,V_{\dot\beta}{}^{\dal}\,,
%\ee
as these annihilate the $(0,1)$-vector fields.
%can be checked by verifying that the above $(0,1)$-vector fields are in the kernel of these 1-forms. 
A computation produces the structure equations $\d e^0=0$ and
\be\label{dtheta}
\d\theta^{\dal} = e^0\wedge\cL_{\p_0}\theta^{\dal} - \theta^{\dot\beta}\wedge\cL_{\p_{\dot\beta}}V^{\dal} - N^{\dal}\,,
\ee
where the ``torsion'' $N^{\dal}$ is found to be
\be\label{N}
%\begin{split}
N\equiv N^{\dal}\,\p_{\dal} = \nbar^2 = \dbar V + \frac{1}{2}\,[V,V]\,.\\
%&= \eps^{\dal\dot\beta}\,\cL_{\p_{\dal}}\!\left(\dbar h + \frac{1}{2}\,\{h,h\}\right)\p_{\dot\beta}\,,
%\end{split}
\ee
This is consistent with the Newlander-Nirenberg theorem that $\nbar^2$ be the obstruction to the integrability of the distribution of $(0,1)$-vector fields. The almost complex structure is integrable precisely when $N$ vanishes.

\medskip

\paragraph{Reconstruction of space-time.} Much like $\R^4$, we can construct the space-time $\cM$ associated to $\CPT$ as a moduli space of rational curves. In homogeneous coordinates, the twistor lines are deformed into a 4-parameter family of pseudo-holomorphic degree 1 rational curves labeled by moduli $y^{\al\dal}$:
\be\label{Ydef}
Y\simeq\P^1 :\;\;\mu^{\dal} =  x^{\al\dal}(y,\lambda)\,\lambda_\al \equiv F^{\dal}(y,\lambda)\,.
\ee
Our space-time $\cM$ is taken to be the moduli space of such curves that are invariant under $Z^A\mapsto\hat Z^A$. It exists and is generically four-dimensional for suitable data $V$ \cite{mcduff2004j,Mason:2005zm}. $\CPT$ is then diffeomorphic to a patch of the projective spinor bundle $\cM\times\P^1\to\cM$ coordinatized by $(y^{\al\dal},\lambda_\al)$.

As the curves are parametrized by $\lambda_\al\in\P^1$, we will abuse notation by using $\{\p_0,\dbar_0\}$ and $\{e^0,\bar e^0\}$ to also denote the standard bases of vector fields and forms on $Y$. In local coordinates, this diffeomorphism can then be expressed as a map
\be
p:\cM\times\P^1\to\CPT,\quad(y,\lambda)\mapsto(x(y,\lambda),\lambda)
\ee
satisfying the PDE for pseudo-holomorphic curves \footnote{The author would like to thank Lionel Mason for pointing out this version of the PDE for the curves.}:
\be\label{f}
\begin{split}
    &\dbar_0\,\lrcorner\,p^*\theta^{\dal} = 0\,,\\
    \text{i.e.,}\quad&\dbar_0F^{\dal} = \dbar_0\,\lrcorner\,p^*V^{\dal}\,.
\end{split}
\ee
When $V=0$, $x^{\al\dal}=y^{\al\dal}$. More generally, one can always solve \eqref{f} for $x^{\al\dal}(y,\lambda)$ locally as it is an elliptic PDE \cite{mcduff2004j}. %This suffices for the purposes of our twistor action.

\medskip

We now prove our new result that constructs an off-shell metric on $\cM$ from the almost complex structure of $\CPT$. The main fact that will be useful for us is that \emph{this does not require complete integrability} $N=0$.
\begin{propn1}[Off-shell Penrose transform]\label{thm1}
$\hspace{1cm}$ Every hamiltonian complex structure deformation $V$ of $\PT$ satisfying $V\,\lrcorner\,e^0=0$ and
\be\label{N1}
\dbar_0\,\lrcorner\,p^*N^{\dal} = 0
\ee
gives rise to a metric on the associated space-time $\cM$.
\end{propn1}

\proof
We begin by introducing a $(2,0)$-form of weight $+2$ in $\lambda_\al$:
\be\label{Sigma}
\Sigma  \vcentcolon= \theta^{\dal}\wedge\theta_{\dal}\,.
\ee
When the almost complex structure is integrable, this gives Gindikin's holomorphic symplectic form on the fibers of $\CPT\to\P^1$ \cite{Gindikin,Mason:2008jy}. %(see \eqref{dSigma} below). 
More generally, it follows from \eqref{dtheta} and \eqref{f} that
\begin{multline}\label{Sigmavar}
\cL_{\dbar_0}p^*(e^0\wedge\Sigma) = \dbar_0\,\lrcorner\,p^*(\text{div}\,V\wedge e^0\wedge \Sigma)\\ + 2\,(\dbar_0\,\lrcorner\,p^*N^{\dal})\wedge\,e^0\wedge p^*\theta_{\dal}\,,
\end{multline}
having noted that $p^*e^0=e^0$ as $p$ preserves $\lambda_\al$. We observe that the first term on the right vanishes due to $V$ being divergence-free, whereas the second term can be made to vanish if and only if we assume \eqref{N1}, i.e., take $N^{\dal}$ to be trivial along the curves.

Hence, assuming that \eqref{divfree} and \eqref{N1} hold, we conclude that $e^0\wedge p^*\Sigma$ is globally holomorphic in $\lambda_\al\in\P^1$. By Liouville's theorem on $\P^1$, we finally obtain a triplet of $2$-forms $\Sigma^{\al\beta} = \Sigma^{(\al\beta)}$ on $\cM$ via
\be\label{Sigmasplit}
p^*\Sigma = \lambda_\al\,\lambda_\beta\,\Sigma^{\al\beta}(y)\qquad\text{mod}\;e^0\,.
\ee
Comparing this with \eqref{Sigma} yields the existence of a matrix $H_{\dot\beta}{}^{\dal}(x,\lambda)\in\SL(2,\C)$ of homogeneity $0$ in $\lambda_\al$ such that
\be\label{thetasplit}
p^*\bigl(H_{\dot\beta}{}^{\dal}\,\theta^{\dot\beta}\bigr) = \lambda_\al\,e^{\al\dal}(y)\qquad\text{mod}\;e^0\,,
\ee
for some 1-forms $e^{\al\dal}$ on $\cM$. In terms of these, we find $\Sigma^{\al\beta} = e^{\al\dal}\wedge e^{\beta}{}_{\dal}$. The $e^{\al\dal}$ comprise a tetrad for our sought after metric \eqref{ds2} on $\cM$. \qed %{\flushright\qed}

\medskip

$H_{\dot\beta}{}^{\dal}(x,\lambda)$ provides a spin-frame on the bundle of dotted spinors $\cO\oplus\cO\to\PS$. Due to being $\SL(2,\C)$-valued, it satisfies $\eps^{\dal\dot\beta}H_{\dal}{}^{\dot\gamma}H_{\dot\beta}{}^{\dot\delta} = \eps^{\dot\gamma\dot\delta}$ and drops out of invariant objects like $\Sigma$. It also solves the d-bar equation
\be\label{dbarH}
\bigl(\delta_{\dot\beta}^{\dot\gamma}\,\dbar_0 + \dbar_0\,\lrcorner\,p^*\cL_{\p_{\dot\beta}}V^{\dot\gamma}\bigr)\,p^*H_{\dot\gamma}{}^{\dal} = 0\,,
\ee
found by acting with $\cL_{\dbar_0}$ on \eqref{thetasplit} and simplifying using \eqref{dtheta}, \eqref{N1}. Alternatively, we can take \eqref{dbarH} as its definition when constructing the action.

As of yet, $e^{\al\dal}$ do not satisfy any equations of motion. The original non-linear graviton construction arises as a corollary of proposition \ref{thm1}. We first use \eqref{divfree} and \eqref{dtheta} to show that
\be\label{dSigma}
\d\Sigma = -2\, N^{\dal}\wedge\theta_{\dal}\qquad\text{mod}\;e^0\,.
\ee
This brings us to
\begin{corol1}[Penrose \cite{Penrose:1976js}]
The resulting metric on $\cM$ is self-dual Ricci-flat if and only if $N^{\dal}=0$.
\end{corol1}
\proof
Pulling back \eqref{dSigma} to $\cM\times\P^1$ using \eqref{Sigmasplit}, we conclude that $\d\Sigma^{\al\beta}(y)=0$ is equivalent to $N^{\dal}=0$. As expected, SD vacuum space-times arise from integrable, hamiltonian complex structure deformations. \qed%{\flushright\qed}

%%%%%%%%%%%%%%%%%%%%%%%
%%%%%%%%%%%%%%%%%%%%%%%

\section{Twistor action for gravity}

Like its space-time counterpart \eqref{plebac}, our proposal for the twistor action decomposes into an action for the SD subsector, plus an interaction term encoding the non-self-dual excitations:
\be\label{mainac}
S[\nbar,B,C] = S_\text{SD}[\nbar,B,C] + \frac{\kappa^2}{4}\,S_\text{int}[\nbar,B]\,.
\ee
This depends on three fields: the almost complex structure represented by the Dolbeault operator $\nbar$ described above, a $(1,1)$-form $B \equiv \theta^{\dal}\wedge B_{\dal}$ with coefficients $B_{\dal}\in\Omega^{0,1}(\PS,\cO(-5))$ acting as a Lagrange multiplier imposing $N\equiv\nbar^2=0$ in the SD subsector, and a further Lagrange multiplier $C\in\Omega^{0,2}(\PS,\cO(-4))$ for the zero-divergence condition \eqref{divfree}. In our correspondence, $\nbar$ is ``dual'' to the space-time metric while $B$ will map to the ASD field $\Gamma_{\al\beta}$ of \eqref{plebac}.

The SD action takes the form
\be\label{sdtac}
S_\text{SD} = \int_{\CPT}\Omega\wedge B_{\dal}\wedge N^{\dal} + \Omega\wedge C\wedge\text{div}\,\theta\,,
\ee
having used the canonical $(3,0)$-form on $\CPT$,
\be
\Omega \vcentcolon=\D\lambda\wedge\Sigma\,,
\ee
of weight $+4$ in $\lambda_\al$. Since $\cL_{\p_{\dal}}e^{\dot\beta}=0$, the equation of motion of $C$ in coordinates is
\be\label{divtheta}
\text{div}\,\theta \equiv \cL_{\p_{\dal}}\theta^{\dal} = -\text{div}\,V = 0\,.
\ee
\eqref{sdtac} is in fact the standard twistor action of self-dual conformal gravity \cite{Mason:2005zm}, now adapted to the fibration $\CPT\to\P^1$ and augmented with the zero-divergence constraint.

On the other hand, the non-self-dual interactions are captured by
\be\label{Iac}
S_\text{int} = \int\displaylimits_{\PS\times_\cM\PS}\la1\,2\ra\,\eps^{\dot\beta_1\dot\beta_2}\bigwedge_{i=1}^2 p_i^*(B_{i\,\dal_i}H_i{}^{\dal_i}{}_{\dot\beta_i}\wedge\Omega_i)
\ee
with the integral being over points $(y^{\al\dal},\lambda_{1\,\al},\lambda_{2\,\al})$ of the fiberwise product $\PS\times_\cM\PS\simeq \cM\times(\P^1)^2$. 
%With a slight abuse of notation, we have extended $f^{-1}:\PS\to\PS$ to the natural map $f^{-1}:\PS\times_\cM\PS\to\PS\times\PS$,
%\be\label{fext}
%f^{-1}:\,(y,\lambda_1,\lambda_2)\mapsto(x(y,\lambda_1),\lambda_1)\times(x(y,\lambda_2),\lambda_2)\,.
%\ee
We have abbreviated $\la\lambda_1\,\lambda_2\ra\equiv\la1\,2\ra$. 
%and defined
%\be
%K_{12}^{\dal\dot\beta}(x,\lambda_1|x,\lambda_2)  \vcentcolon= \frac{H^{\dal\dot\gamma}(x,\lambda_1)\,H^{\dot\beta}{}_{\dot\gamma}(x,\lambda_2)}{\la1\,2\ra}
%\ee
%which is a Green's function for the PDE \eqref{dbarH}. 
The remaining ingredients are $B_i \equiv B(x,\lambda_i)$, $\Omega_i \equiv \Omega(x,\lambda_i)$, $H_i{}^{\dal}{}_{\dot\beta}\equiv H^{\dal}{}_{\dot\beta}(x,\lambda_i)$, so that $p_i^*B_i \equiv B_i(x(y,\lambda_i),\lambda_i)$, etc. This interaction is non-local on twistor space but reduces to the expected interaction term on space-time. 
%However, expressing \eqref{Iac} in terms of the fundamental fields $h$ and $\tilde h$ off-shell is still an unresolved issue.

As written, our twistor action is only partially covariant as it singles out the coordinate $\lambda_\al$ along the base of the fibration $\CPT\to\P^1$. It is only invariant under diffeomorphisms preserving the fibration and the Poisson bivector in \eqref{I}, as these also preserve the zero-divergence constraint. Nevertheless, this is enough to give rise to the covariant action \eqref{plebac} on space-time.

%%%%%%%%%%%%%%%%%%%%%%%
%%%%%%%%%%%%%%%%%%%%%%%

\section{Equivalence with GR}

Recall the diffeomorphism $p:\cM\times\P^1\to\PS\simeq\CPT$. We now prove that we can compactify the twistor action \eqref{mainac} along $\P^1$ to obtain the \emph{complete} action \eqref{plebac} of GR as the effective theory on $\cM$.

\begin{propn1}\label{prop2}
On performing the Penrose transform,
\begin{enumerate}[(i)]
    \item the SD twistor action \eqref{sdtac} reduces to the SD space-time action \eqref{sdplebac}.
    \item The non-SD interaction term \eqref{Iac} reduces to the interaction term in the space-time action \eqref{plebac}.
\end{enumerate}
\end{propn1}

\proof
\textbf{(\textit{i})} Given a $V$, we will integrate out the components of $B$ orthogonal to its pseudo-holomorphic curves. Using $\{e^0,\bar e^0, \d y^{\al\dal}\}$ as a basis of $T^*(\cM\times\P^1)$, we split %\footnote{As this splitting depends on $V$, it may introduce interesting quantum anomalies in the $V$ path integral. We neglect these in our classical treatment.}
\be\label{Bdec}
p^*B_{\dal} = B_{\dal}\bigr|_Y + B_{\dal}^\perp\,,
\ee
where $B_{\dal}|_Y$ is in the span of $e^0,\bar e^0$ and $B_{\dal}^\perp$ in the span of $\d y^{\al\dal}$. Next note that due to \eqref{f}, $\dbar_0\,\lrcorner\,p^*\Omega=0$, i.e., $p^*\Omega$ has no $\bar e^0$-component. So, the integral in \eqref{Iac} can only be saturated by the $\bar e^0$-component of $p^*B_{\dal}$. Consequently, $B^\perp_{\dal}$ drops out of $S_\text{int}$. It will only be present in $S_\text{SD}$.

We begin the reduction by integrating out $C$. This imposes \eqref{divtheta} as a constraint. In the remaining SD action, we perform a ``change of integration variables'' using the diffeomorphism $p$. This yields
\be\label{compsdac}
\begin{split}
    S_\text{SD} &= \int_{\cM\times\P^1}p^*(\Omega\wedge B_{\dal}\wedge N^{\dal})\\
    &= \int_{\cM\times Y}p^*\Omega\wedge\left(B_{\dal}\bigr|_Y + B_{\dal}^\perp\right)\wedge p^*N^{\dal}\,,
\end{split}
\ee
having used \eqref{Bdec}. Again, since $\dbar_0\,\lrcorner\,p^*\Omega = 0 = \dbar_0\,\lrcorner\,B_{\dal}^\perp$, the second term in this integral only retains the component of $p^*N^{\dal}$ along $\bar e_0$. Thus, integrating out $B_{\dal}^\perp$ from the theory %from \eqref{compsdac} (or more precisely $(p^{-1})^*B_{\dal}^\perp$ from \eqref{mainac}) 
imposes $\dbar_0\,\lrcorner\,p^*N^{\dal}=0$ as a constraint. 

By proposition \ref{thm1}, this allows us to construct a space-time tetrad $e^{\al\dal}$ as in \eqref{thetasplit}. With this in hand, we can use $\Omega = \D\lambda\wedge\theta^{\dal}\wedge\theta_{\dal}$ to recast the rest of the SD action as
\begin{align}
    &S_\text{SD} = -2\int_{\cM\times Y} \D\lambda\wedge p^*\theta^{\dal}\wedge B_{\dal}\bigr|_Y\wedge p^*(N^{\dot\beta}\wedge\theta_{\dot\beta})\nonumber\\
    &= \int_{\cM\times Y} \D\lambda\wedge p^*\theta^{\dal}\wedge B_{\dal}\bigr|_Y\wedge p^*\d\Sigma\\
    &= \int_{\cM\times Y} \D\lambda\wedge p^*\theta^{\dal}\wedge B_{\dal}\bigr|_Y\wedge\lambda_\al\,\lambda_\beta\,\d\Sigma^{\al\beta}(y)\,.\nonumber
\end{align}
To get the first line, we have used $2\,\theta^{\dal}\wedge\theta_{\dot\beta} = \delta^{\dal}_{\dot\beta}\,\theta^{\dot\gamma}\wedge\theta_{\dot\gamma}$. The second line follows from \eqref{dSigma}, while the third line is a consequence of \eqref{Sigmasplit}. Hence, defining the space-time field $\Gamma_{\al\beta}(y)$ as the Penrose transform of $B$ \cite{Mason:2008jy},
\be\label{GammaPenrose}
\begin{split}
\Gamma_{\al\beta}(y) &\vcentcolon= \int_{Y}\D\lambda\wedge\lambda_\al\,\lambda_\beta\,p^*B\\
&= \int_{Y}\D\lambda\wedge\lambda_\al\,\lambda_\beta\, p^*\theta^{\dal}\wedge B_{\dal}\bigr|_Y\,,
\end{split}
\ee
the SD action reduces to \eqref{sdplebac}.

\medskip

\textbf{(\textit{ii})} Using \eqref{GammaPenrose}, we can also prove the equivalence of the interaction term in \eqref{plebac} with the non-SD action \eqref{Iac}:
\begin{align}
    &\;\kappa^2\int_\cM\Sigma^{\al\beta}\wedge\Gamma_\al{}^\gamma\wedge\Gamma_{\gamma\beta}\nonumber\\
    =&\; \kappa^2\,\int_{\PS\times_\cM\PS}\la1\,2\ra\,\lambda_{1\,\al}\,e^{\al\dal}\wedge\lambda_{2\,\beta}\,e^{\beta}{}_{\dal}\nonumber\\
    &\quad\wedge\D\lambda_1\wedge p_1^*\bigl(\theta_1^{\dot\beta}\wedge B_{1\,\dot\beta}\bigr)\wedge\D\lambda_2\wedge p_2^*\bigl(\theta_2^{\dot\gamma}\wedge B_{2\,\dot\gamma}\bigr)\nonumber\\
    =&\, \frac{\kappa^2}{4}\,S_\text{int}\,,
\end{align}
with $\theta_i^{\dal}\equiv\theta^{\dal}(x,\lambda_i)$, etc. To get the second equality, we have substituted for $\lambda_{i\,\al}\,e^{\al\dal}$ using \eqref{thetasplit}.\qed
%{\flushright\qed}
\medskip

\eqref{GammaPenrose} is invariant under $B\mapsto B+\nbar\chi$ for smooth $(1,0)$-forms $\chi = \chi_{\dal}(x,\lambda)\,\theta^{\dal}\in\Omega^{1,0}(\CPT,\cO(-4))$. This gives an additional gauge symmetry of the twistor action. The invariance of $S_\text{int}$ holds by construction, while that of $S_\text{SD}$ follows from symmetries of the SD conformal gravity twistor action \cite{Mason:2005zm} and because $\nbar\Omega\wedge\chi_{\dal}N^{\dal}\propto\text{div}\,V=0$. This extra gauge symmetry is expected to be a crucial ingredient for the derivation of gravitational MHV rules, as it allows for useful gauge choices \cite{Cachazo:2004kj,Boels:2007qn,Mason:2008jy,Adamo:2013tja}.

%%%%%%%%%%%%%%%%%%%%%%%
%%%%%%%%%%%%%%%%%%%%%%%

\section{Discussion}

We have proposed a new twistor action for gravity. Proposition \ref{prop2} shows that it is equivalent to Plebanski's formulation of Euclidean general relativity. Solutions to its field equations should yield twistor spaces for generic vacuum space-times that are not necessarily self-dual, while its perturbation theory should provide new principles for the computation of graviton amplitudes.

Previous attempts like \cite{Herfray:2016qvg} built covariant twistor actions by means of the Atiyah-Hitchin-Singer almost complex structure, but were unable to reduce them to GR by compactification on $\P^1$. %In particular, they lacked a covariant analogue of our zero-divergence constraint. 
Unlike them, we have built an analogue of Kodaira-Spencer gravity \cite{Bershadsky:1993cx} by taking the almost complex structure of $\CPT$ to be a deformation of the complex structure of the ``flat'' background $\PT$. We also assumed the existence of a smooth fibration $\CPT\to\P^1$, as this appropriately reduced the degrees of freedom in compatible deformations $V$. An important next step would be to complete this into a fully covariant, background-independent formalism, along with studying its quantum consistency.

A primary motivation for our construction is to derive robust gravitational MHV rules from an action principle. A preliminary study of the MHV vertices originating from \eqref{Iac} has already been performed by using them to compute on-shell graviton MHV amplitudes in flat space \cite{Mason:2008jy, Adamo:2021bej} as well as in non-trivial classes of self-dual space-times \cite{Adamo:2020syc}. It would be very interesting to see if the methods of these works can be applied to the off-shell MHV vertices and propagator of our action. 

It should also be possible to adapt our twistor action to the case of a non-vanishing cosmological constant as well as to non-trivial amounts of supersymmetries. This could give insights into MHV rules for graviton scattering in (A)dS$_4$, possibly making contact with the worldsheet formulae in \cite{Adamo:2015ina}. Our action may also find analogues on the spinor bundles of Lorentzian or split-signature space-times \cite{Jiang:2008xw,Bittleston:2020hfv}, in other dimensions \cite{Adamo:2017xaf}, holographic backgrounds \cite{Adamo:2016rtr}, and integrable systems \cite{Bittleston:2020hfv, Penna:2020uky}.

It would also be interesting to explore the recently discovered twistorial origins of double copy \cite{White:2020sfn,Chacon:2021wbr} at the level of twistor actions. This could help extend the analysis of the kinematic algebra in \cite{Monteiro:2011pc} to the non-self-dual sector. Further points of interest include building connections with other contemporary work on reformulations of GR. See for instance  \cite{Krasnov:2020bqr} for graviton scattering using the chiral action on space-time, and \cite{Ashtekar:2020xll} for possible implications for double copy. It would also be worthwhile to try developing links with inherently quantum formulations like loop quantum gravity \cite{Speziale:2012nu}.

%%%%%%%%%%%%%%%%%%%%%%%
%%%%%%%%%%%%%%%%%%%%%%%

\medskip

\textit{Acknowledgments:} It is a pleasure to thank Tim Adamo, Roland Bittleston, Lionel Mason and David Skinner for helpful discussions and comments on the draft. AS is supported by a Mathematical Institute Studentship, Oxford.

\bibliographystyle{JHEP}
\bibliography{gta}

\providecommand{\href}[2]{#2}\begingroup\raggedright\begin{thebibliography}{10}

\bibitem{Witten:2003nn}
E.~Witten, {\it {Perturbative gauge theory as a string theory in twistor
  space}},  {\em Commun. Math. Phys.} {\bf 252} (2004) 189--258,
  [\href{http://arxiv.org/abs/hep-th/0312171}{{\tt hep-th/0312171}}].

\bibitem{Berkovits:2004hg}
N.~Berkovits, {\it {An Alternative string theory in twistor space for N=4
  superYang-Mills}},  {\em Phys. Rev. Lett.} {\bf 93} (2004) 011601,
  [\href{http://arxiv.org/abs/hep-th/0402045}{{\tt hep-th/0402045}}].

\bibitem{Roiban:2004yf}
R.~Roiban, M.~Spradlin, and A.~Volovich, {\it {On the tree level S matrix of
  Yang-Mills theory}},  {\em Phys. Rev. D} {\bf 70} (2004) 026009,
  [\href{http://arxiv.org/abs/hep-th/0403190}{{\tt hep-th/0403190}}].

\bibitem{Geyer:2014fka}
Y.~Geyer, A.~E. Lipstein, and L.~J. Mason, {\it {Ambitwistor Strings in Four
  Dimensions}},  {\em Phys. Rev. Lett.} {\bf 113} (2014), no.~8 081602,
  [\href{http://arxiv.org/abs/1404.6219}{{\tt arXiv:1404.6219}}].

\bibitem{Mason:2005zm}
L.~J. Mason, {\it {Twistor actions for non-self-dual fields: A Derivation of
  twistor-string theory}},  {\em JHEP} {\bf 10} (2005) 009,
  [\href{http://arxiv.org/abs/hep-th/0507269}{{\tt hep-th/0507269}}].

\bibitem{Boels:2006ir}
R.~Boels, L.~J. Mason, and D.~Skinner, {\it {Supersymmetric Gauge Theories in
  Twistor Space}},  {\em JHEP} {\bf 02} (2007) 014,
  [\href{http://arxiv.org/abs/hep-th/0604040}{{\tt hep-th/0604040}}].

\bibitem{Cachazo:2004kj}
F.~Cachazo, P.~Svrcek, and E.~Witten, {\it {MHV vertices and tree amplitudes in
  gauge theory}},  {\em JHEP} {\bf 09} (2004) 006,
  [\href{http://arxiv.org/abs/hep-th/0403047}{{\tt hep-th/0403047}}].

\bibitem{Boels:2007qn}
R.~Boels, L.~J. Mason, and D.~Skinner, {\it {From twistor actions to MHV
  diagrams}},  {\em Phys. Lett. B} {\bf 648} (2007) 90--96,
  [\href{http://arxiv.org/abs/hep-th/0702035}{{\tt hep-th/0702035}}].

\bibitem{Adamo:2011cb}
T.~Adamo and L.~Mason, {\it {MHV diagrams in twistor space and the twistor
  action}},  {\em Phys. Rev. D} {\bf 86} (2012) 065019,
  [\href{http://arxiv.org/abs/1103.1352}{{\tt arXiv:1103.1352}}].

\bibitem{Mason:2010yk}
L.~J. Mason and D.~Skinner, {\it {The Complete Planar S-matrix of N=4 SYM as a
  Wilson Loop in Twistor Space}},  {\em JHEP} {\bf 12} (2010) 018,
  [\href{http://arxiv.org/abs/1009.2225}{{\tt arXiv:1009.2225}}].

\bibitem{Bullimore:2011ni}
M.~Bullimore and D.~Skinner, {\it {Holomorphic Linking, Loop Equations and
  Scattering Amplitudes in Twistor Space}},
  \href{http://arxiv.org/abs/1101.1329}{{\tt arXiv:1101.1329}}.

\bibitem{Adamo:2011pv}
T.~Adamo, M.~Bullimore, L.~Mason, and D.~Skinner, {\it {Scattering Amplitudes
  and Wilson Loops in Twistor Space}},  {\em J. Phys. A} {\bf 44} (2011)
  454008, [\href{http://arxiv.org/abs/1104.2890}{{\tt arXiv:1104.2890}}].

\bibitem{Adamo:2011dq}
T.~Adamo, M.~Bullimore, L.~Mason, and D.~Skinner, {\it {A Proof of the
  Supersymmetric Correlation Function / Wilson Loop Correspondence}},  {\em
  JHEP} {\bf 08} (2011) 076, [\href{http://arxiv.org/abs/1103.4119}{{\tt
  arXiv:1103.4119}}].

\bibitem{Adamo:2011cd}
T.~Adamo, {\it {Correlation functions, null polygonal Wilson loops, and local
  operators}},  {\em JHEP} {\bf 12} (2011) 006,
  [\href{http://arxiv.org/abs/1110.3925}{{\tt arXiv:1110.3925}}].

\bibitem{Koster:2014fva}
L.~Koster, V.~Mitev, and M.~Staudacher, {\it {A Twistorial Approach to
  Integrability in $\mathcal N=$ 4 SYM}},  {\em Fortsch. Phys.} {\bf 63}
  (2015), no.~2 142--147, [\href{http://arxiv.org/abs/1410.6310}{{\tt
  arXiv:1410.6310}}].

\bibitem{Chicherin:2014uca}
D.~Chicherin, R.~Doobary, B.~Eden, P.~Heslop, G.~P. Korchemsky, L.~Mason, and
  E.~Sokatchev, {\it {Correlation functions of the chiral stress-tensor
  multiplet in $ \mathcal{N}=4 $ SYM}},  {\em JHEP} {\bf 06} (2015) 198,
  [\href{http://arxiv.org/abs/1412.8718}{{\tt arXiv:1412.8718}}].

\bibitem{Koster:2016ebi}
L.~Koster, V.~Mitev, M.~Staudacher, and M.~Wilhelm, {\it {Composite Operators
  in the Twistor Formulation of N=4 Supersymmetric Yang-Mills Theory}},  {\em
  Phys. Rev. Lett.} {\bf 117} (2016), no.~1 011601,
  [\href{http://arxiv.org/abs/1603.04471}{{\tt arXiv:1603.04471}}].

\bibitem{Koster:2016loo}
L.~Koster, V.~Mitev, M.~Staudacher, and M.~Wilhelm, {\it {All tree-level MHV
  form factors in $ \mathcal{N} $ = 4 SYM from twistor space}},  {\em JHEP}
  {\bf 06} (2016) 162, [\href{http://arxiv.org/abs/1604.00012}{{\tt
  arXiv:1604.00012}}].

\bibitem{Koster:2016fna}
L.~Koster, V.~Mitev, M.~Staudacher, and M.~Wilhelm, {\it {On Form Factors and
  Correlation Functions in Twistor Space}},  {\em JHEP} {\bf 03} (2017) 131,
  [\href{http://arxiv.org/abs/1611.08599}{{\tt arXiv:1611.08599}}].

\bibitem{Adamo:2020syc}
T.~Adamo, L.~Mason, and A.~Sharma, {\it {MHV scattering of gluons and gravitons
  in chiral strong fields}},  {\em Phys. Rev. Lett.} {\bf 125} (2020), no.~4
  041602, [\href{http://arxiv.org/abs/2003.13501}{{\tt arXiv:2003.13501}}].

\bibitem{Adamo:2020yzi}
T.~Adamo, L.~Mason, and A.~Sharma, {\it {Gluon scattering on self-dual
  radiative gauge fields}},  \href{http://arxiv.org/abs/2010.14996}{{\tt
  arXiv:2010.14996}}.

\bibitem{Skinner:2013xp}
D.~Skinner, {\it {Twistor strings for $ \mathcal{N} $ = 8 supergravity}},  {\em
  JHEP} {\bf 04} (2020) 047, [\href{http://arxiv.org/abs/1301.0868}{{\tt
  arXiv:1301.0868}}].

\bibitem{Hodges:2012ym}
A.~Hodges, {\it {A simple formula for gravitational MHV amplitudes}},
  \href{http://arxiv.org/abs/1204.1930}{{\tt arXiv:1204.1930}}.

\bibitem{Cachazo:2012kg}
F.~Cachazo and D.~Skinner, {\it {Gravity from Rational Curves in Twistor
  Space}},  {\em Phys. Rev. Lett.} {\bf 110} (2013), no.~16 161301,
  [\href{http://arxiv.org/abs/1207.0741}{{\tt arXiv:1207.0741}}].

\bibitem{Cachazo:2012pz}
F.~Cachazo, L.~Mason, and D.~Skinner, {\it {Gravity in Twistor Space and its
  Grassmannian Formulation}},  {\em SIGMA} {\bf 10} (2014) 051,
  [\href{http://arxiv.org/abs/1207.4712}{{\tt arXiv:1207.4712}}].

\bibitem{BjerrumBohr:2005jr}
N.~E.~J. Bjerrum-Bohr, D.~C. Dunbar, H.~Ita, W.~B. Perkins, and K.~Risager,
  {\it {MHV-vertices for gravity amplitudes}},  {\em JHEP} {\bf 01} (2006) 009,
  [\href{http://arxiv.org/abs/hep-th/0509016}{{\tt hep-th/0509016}}].

\bibitem{Bianchi:2008pu}
M.~Bianchi, H.~Elvang, and D.~Z. Freedman, {\it {Generating Tree Amplitudes in
  N=4 SYM and N = 8 SG}},  {\em JHEP} {\bf 09} (2008) 063,
  [\href{http://arxiv.org/abs/0805.0757}{{\tt arXiv:0805.0757}}].

\bibitem{Berkovits:2004jj}
N.~Berkovits and E.~Witten, {\it {Conformal supergravity in twistor-string
  theory}},  {\em JHEP} {\bf 08} (2004) 009,
  [\href{http://arxiv.org/abs/hep-th/0406051}{{\tt hep-th/0406051}}].

\bibitem{Wolf:2007tx}
M.~Wolf, {\it {Self-Dual Supergravity and Twistor Theory}},  {\em Class. Quant.
  Grav.} {\bf 24} (2007) 6287--6328,
  [\href{http://arxiv.org/abs/0705.1422}{{\tt arXiv:0705.1422}}].

\bibitem{Mason:2007ct}
L.~J. Mason and M.~Wolf, {\it {Twistor Actions for Self-Dual Supergravities}},
  {\em Commun. Math. Phys.} {\bf 288} (2009) 97--123,
  [\href{http://arxiv.org/abs/0706.1941}{{\tt arXiv:0706.1941}}].

\bibitem{Mason:2008jy}
L.~J. Mason and D.~Skinner, {\it {Gravity, Twistors and the MHV Formalism}},
  {\em Commun. Math. Phys.} {\bf 294} (2010) 827--862,
  [\href{http://arxiv.org/abs/0808.3907}{{\tt arXiv:0808.3907}}].

\bibitem{Adamo:2013tja}
T.~Adamo and L.~Mason, {\it {Conformal and Einstein gravity from twistor
  actions}},  {\em Class. Quant. Grav.} {\bf 31} (2014), no.~4 045014,
  [\href{http://arxiv.org/abs/1307.5043}{{\tt arXiv:1307.5043}}].

\bibitem{Adamo:2013cra}
T.~Adamo, {\em {Twistor actions for gauge theory and gravity}}.
\newblock PhD thesis, University of Oxford, 2013.
\newblock \href{http://arxiv.org/abs/1308.2820}{{\tt arXiv:1308.2820}}.

\bibitem{Adamo:2021bej}
T.~Adamo, L.~Mason, and A.~Sharma, {\it {Twistor sigma models for quaternionic
  geometry and graviton scattering}},
  \href{http://arxiv.org/abs/2103.16984}{{\tt arXiv:2103.16984}}.

\bibitem{Herfray:2016qvg}
Y.~Herfray, {\it {Pure Connection Formulation, Twistors and the Chase for a
  Twistor Action for General Relativity}},  {\em J. Math. Phys.} {\bf 58}
  (2017), no.~11 112505, [\href{http://arxiv.org/abs/1610.02343}{{\tt
  arXiv:1610.02343}}].

\bibitem{AbouZeid:2005dg}
M.~Abou-Zeid and C.~M. Hull, {\it {A Chiral perturbation expansion for
  gravity}},  {\em JHEP} {\bf 02} (2006) 057,
  [\href{http://arxiv.org/abs/hep-th/0511189}{{\tt hep-th/0511189}}].

\bibitem{Penrose:1976js}
R.~Penrose, {\it {Nonlinear Gravitons and Curved Twistor Theory}},  {\em Gen.
  Rel. Grav.} {\bf 7} (1976) 31--52.

\bibitem{Penrose:2015lla}
R.~Penrose, {\it {Palatial twistor theory and the twistor googly problem}},
  {\em Phil. Trans. Roy. Soc. Lond. A} {\bf 373} (2015) 20140237.

\bibitem{Note1}
We will neglect the boundary term as we are restricting attention to
  asymptotically flat space-times. Like the topological term, it may need to be
  reinstated when turning on a cosmological constant.

\bibitem{Woodhouse:1985id}
N.~M.~J. Woodhouse, {\it {Real methods in twistor theory}},  {\em Class. Quant.
  Grav.} {\bf 2} (1985) 257--291.

\bibitem{Jiang:2008xw}
W.~Jiang, {\em {Aspects of Yang-Mills Theory in Twistor Space}}.
\newblock PhD thesis, University of Oxford, 2008.
\newblock \href{http://arxiv.org/abs/0809.0328}{{\tt arXiv:0809.0328}}.

\bibitem{Adamo:2017qyl}
T.~Adamo, {\it {Lectures on twistor theory}},  {\em PoS} {\bf Modave2017}
  (2018) 003, [\href{http://arxiv.org/abs/1712.02196}{{\tt arXiv:1712.02196}}].

\bibitem{Atiyah:1978wi}
M.~F. Atiyah, N.~J. Hitchin, and I.~M. Singer, {\it {Selfduality in
  Four-Dimensional Riemannian Geometry}},  {\em Proc. Roy. Soc. Lond. A} {\bf
  362} (1978) 425--461.

\bibitem{Ward:1990vs}
R.~S. Ward and R.~O. Wells, {\em {Twistor geometry and field theory}}.
\newblock Cambridge Monographs on Mathematical Physics. Cambridge University
  Press, 8, 1991.

\bibitem{mcduff2004j}
D.~McDuff and D.~Salamon, {\it J-holomorphic curves and symplectic topology},
  in {\em Colloquium Publications}, vol.~52, AMS, 2004.

\bibitem{Note2}
The author would like to thank Lionel Mason for pointing out this version of
  the PDE for the curves.

\bibitem{Gindikin}
S.~G. Gindikin, {\it {A construction of hyper-K\"ahler metrics}},  {\em Funct.
  Anal. Appl.} {\bf 20} (1986) 238–240.

\bibitem{Bershadsky:1993cx}
M.~Bershadsky, S.~Cecotti, H.~Ooguri, and C.~Vafa, {\it {Kodaira-Spencer theory
  of gravity and exact results for quantum string amplitudes}},  {\em Commun.
  Math. Phys.} {\bf 165} (1994) 311--428,
  [\href{http://arxiv.org/abs/hep-th/9309140}{{\tt hep-th/9309140}}].

\bibitem{Adamo:2015ina}
T.~Adamo, {\it {Gravity with a cosmological constant from rational curves}},
  {\em JHEP} {\bf 11} (2015) 098, [\href{http://arxiv.org/abs/1508.02554}{{\tt
  arXiv:1508.02554}}].

\bibitem{Bittleston:2020hfv}
R.~Bittleston and D.~Skinner, {\it {Twistors, the ASD Yang-Mills equations, and
  4d Chern-Simons theory}},  \href{http://arxiv.org/abs/2011.04638}{{\tt
  arXiv:2011.04638}}.

\bibitem{Adamo:2017xaf}
T.~Adamo, D.~Skinner, and J.~Williams, {\it {Minitwistors and 3d
  Yang-Mills-Higgs theory}},  {\em J. Math. Phys.} {\bf 59} (2018), no.~12
  122301, [\href{http://arxiv.org/abs/1712.09604}{{\tt arXiv:1712.09604}}].

\bibitem{Adamo:2016rtr}
T.~Adamo, D.~Skinner, and J.~Williams, {\it {Twistor methods for AdS$_{5}$}},
  {\em JHEP} {\bf 08} (2016) 167, [\href{http://arxiv.org/abs/1607.03763}{{\tt
  arXiv:1607.03763}}].

\bibitem{Penna:2020uky}
R.~F. Penna, {\it {A Twistor Action for Integrable Systems}},
  \href{http://arxiv.org/abs/2011.05831}{{\tt arXiv:2011.05831}}.

\bibitem{White:2020sfn}
C.~D. White, {\it {Twistorial Foundation for the Classical Double Copy}},  {\em
  Phys. Rev. Lett.} {\bf 126} (2021), no.~6 061602,
  [\href{http://arxiv.org/abs/2012.02479}{{\tt arXiv:2012.02479}}].

\bibitem{Chacon:2021wbr}
E.~Chac\'on, S.~Nagy, and C.~D. White, {\it {The Weyl double copy from twistor
  space}},  \href{http://arxiv.org/abs/2103.16441}{{\tt arXiv:2103.16441}}.

\bibitem{Monteiro:2011pc}
R.~Monteiro and D.~O'Connell, {\it {The Kinematic Algebra From the Self-Dual
  Sector}},  {\em JHEP} {\bf 07} (2011) 007,
  [\href{http://arxiv.org/abs/1105.2565}{{\tt arXiv:1105.2565}}].

\bibitem{Krasnov:2020bqr}
K.~Krasnov and Y.~Shtanov, {\it {Chiral perturbation theory for GR}},  {\em
  JHEP} {\bf 09} (2020) 017, [\href{http://arxiv.org/abs/2007.00995}{{\tt
  arXiv:2007.00995}}].

\bibitem{Ashtekar:2020xll}
A.~Ashtekar and M.~Varadarajan, {\it {Gravitational Dynamics\textemdash{}A
  Novel Shift in the Hamiltonian Paradigm}},  {\em Universe} {\bf 7} (2021),
  no.~1 13, [\href{http://arxiv.org/abs/2012.12094}{{\tt arXiv:2012.12094}}].

\bibitem{Speziale:2012nu}
S.~Speziale and W.~M. Wieland, {\it {The twistorial structure of loop-gravity
  transition amplitudes}},  {\em Phys. Rev. D} {\bf 86} (2012) 124023,
  [\href{http://arxiv.org/abs/1207.6348}{{\tt arXiv:1207.6348}}].

\end{thebibliography}\endgroup

\end{document}